\documentclass{article}
\usepackage{graphicx} 
\usepackage[sorting=none]{biblatex}
\usepackage{amsmath}
\usepackage[table]{xcolor}
\usepackage{comment}

\title{Hash chaining degrades security at Facebook}
\author{Thomas Rivasseau\\
\textit{McGill University}}
\date{October 2025}

\begin{document}

\maketitle

\begin{abstract}
    Modern web and digital application password storage relies on password hashing for storage and security. Ad-hoc upgrade of password storage to keep up with hash algorithm norms may be used to save costs but can introduce unforeseen vulnerabilities. This is the case in the password storage scheme used by Meta Platforms which services several billion monthly users worldwide. In this paper we present the first example of an exploit which demonstrates the security weakness of Facebook's password storage scheme, and discuss its implications. Proper ethical disclosure guidelines and vendor notification were followed.
\end{abstract}

\section{Introduction}
Password hashing is the standard way to store user passwords in web and other digital applications. This provides security in case the user password database is leaked, which is frequent \cite{guardian}. Security of password hashing relies on several elements but mainly the underlying hash algorithm. We will provide an overview of password hashing, how it is used for security applications, and showcase the password hashing scheme used to secure user accounts at Facebook, or Meta Platforms Inc. We then show that this storage system presents a vulnerability which makes it significantly less secure than anticipated. We also provide an exploit which demonstrates our vulnerability. Then we explain the consequences of this exploit, and limitations. We conclude and provide ethical and security disclaimers.

\section{Password hashing}
In this section, we provide an overview of hashing algorithms and their use in modern account password storage infrastructure. Hashing algorithms are usually traced back to the 1950's \cite{hash_birth}. They have become a practical way to compress data for fast access, analysis, and verification \cite{hashing_explained}. The NIST Computer Security Resource Center defines hash functions as \cite{nist}:
\begin{center}
    A function on bit strings in which the length of the output is fixed. Approved hash functions (such as those specified in FIPS 180 and FIPS 202) are designed to satisfy the following properties:

1. (One-way) It is computationally infeasible to find any input that maps to any new pre-specified output.

2. (Collision-resistant) It is computationally infeasible to find any two distinct inputs that map to the same output.
\end{center}
Hash functions are complex algorithms that output a value for a given input. This value is linked to the input, so an input will always produce the same output when hashed. The output cannot be traced back to the input, and this is one-wayness. Also, outputs appear random, and it should be impossible to find two inputs that have the same output. This is collision resistance.\\
Hash functions are useful for digital signatures and data authentication because if someone knows the hash of a digital document, they can verify if a copy of said document has been altered. This is because any change to a document will produce a hash that is different from the original. As specified in the official recommendations of Canada \cite{CAN_CSRC}, only secure hashing algorithms should be employed. Specifically, SHA-256 and above or SHA3-256 and above. Here, "above" means with greater output length. A hash function's output length is specified in bits and describes how long the output of the function is. It is recommended to have at least a 256-bit output length, which corresponds to 32 characters. Larger output lengths mean more possible outputs, which strengthens security.\\
Hash functions are used for password storage. According to OWASP, passwords "must be protected from an attacker even if the application or database is compromised" \cite{OWASP}. Password database leaks are frequent \cite{guardian}. Thus, websites only store the hash value of a password, not the password itself. It is easy to compute the hash of what the user submits as password to login, and compare it with the value stored. If both values match, hash collision-resistance means both inputs were identical, so the user has submitted the correct password \\ 
If an attacker obtains the value stored at the server, they cannot find the password, and cannot login. Additionally, it is not feasible to find a wrong password that generates the correct hash value. \\
Once hash value is leaked, attackers will attempt to guess the password. This is called brute force, or exhaustive search \cite{Brute_f}. Attackers will compute the hash of all possible passwords and compare it with the leaked value. If it matches, they have guessed the password. This technique can be optimized by guessing popular passwords first like "00000", "12345" or "hello". Databases exist of the most common 1000, 10000 and even 1 Million passwords \cite{pw_db}. \\ 
Password hashing can be made even safer by salting \cite{salting}. Salting means adding a random value to the password before computing the hash. The value is stored as-is next to the hash. This prevents an attacker from re-using their previous password guesses. Re-using password guesses is a dictionary attack \cite{rainbow}.\\
Another technique to prevent brute force attacks is to increase the time needed to compute the hash value from a password. It is done by computing the hash value sequentially many times \cite{PKBDF,OWASP}.\\
In this section we have provided a brief overview of hashing and how it is used to store passwords in modern digital applications and websites.

\section{Case study: Facebook}
\begin{figure}
\begin{center}
  \includegraphics[width = \linewidth]{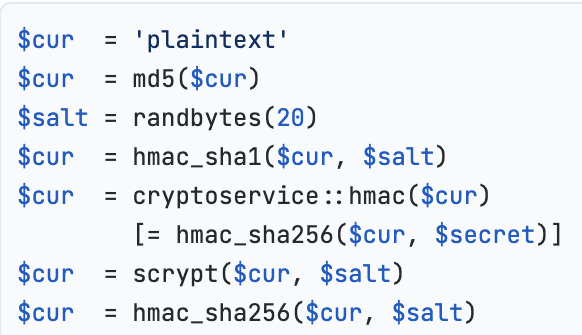}
  \caption{Facebook's password hashing scheme as presented in 2014}
  \label{fb_2014}
  \end{center}
\end{figure}

In this section, we will present the password storage system of a well-known technology provider: Meta Platforms Inc., particularly the Facebook website. The company boasts about 4 billion monthly active users \cite{fb_users} across its products, is the leading social media brand worldwide and is valued at 1.8 trillion USD \cite{fb_mc}.
In december 2014, Alec Muffett and Andrei Bajenov, both engineers working at Facebook explained at the 7th International Conference on Passwords in Trondheim how their company stores its user passwords \cite{Passresearch}. The recording of this talk can be found on Youtube \cite{fb_pw_h}. Should the video prove hard to find, several blogs also explain its contents \cite{crypto_fb,Bristol} and that of an identical talk \cite{RWC_2015}.
At the time, Facebook was growing rapidly with a user base approaching the billion, and sought an appropriate way to upgrade its password storage systems. Early 2000s password storage did not use salting or memory hardness and was done using a hashing algorithm called MD5 \cite{MD5}. MD5 collision attacks were demonstrated as early as 2005 when a team led by Arjen Lenstra crafted two valid X.509 digital certificates with identical signatures \cite{lenstra}. In 2008, the Software institute at Carnegie Mellon University issued a recommendation to no longer use MD5 for security applications \cite{CMU}.\\
For Facebook engineers, the problem of migrating from MD5 to a safer algorithm was that it would be costly and imply keeping two hash tables. One old table with the MD5 hashes and one new, with the more modern SHA1 hashes. Every time a user logged in, their password would be checked with the MD5 value to make sure it was correct. Then the SHA1 hash would be calculated and stored, and when all users did this, the old MD5 hashes could be deleted \cite{Bristol}. This would be expensive, long, and pose problems with users who would not be logging in for prolonged periods.\\
Instead, Facebook engineers hashed their MD5 password hash values, not the user passwords, with SHA1. Password storage thus became a chain of two hash algorithms: MD5 followed by SHA1. When users logged in, their candidate password was first hashed with MD5, then SHA1, then compared to the value in memory and if identical, login was authorized. This removed the need for costly system overhaul.\\
Then Facebook did this again to add salts and a memory hard function, and upgrade the hash algorithm once more to SHA256. At every stage they lengthened the layered hash chain, and kept only one password hash value in storage, avoiding database upgrade costs.\\
A screenshot of the final system architecture is shown in Fig \ref{fb_2014}\cite{crypto_fb,fb_pw_h}.
The password ("plaintext") is first hashed using MD5. As explained previously this is because all passwords stored before the initial update were in MD5 hash format. A salt is then initiated and added to a SHA1 hash of the previous value.\\
With old passwords, this is done by hashing the value stored at the server with SHA1. New passwords are put through the full hash chain.\\
Similar hashing logic is applied further down the line. The SHA1 output is fed to the more secure SHA256, with a different salt, which may be stored separately. The output is fed to scrypt which is a memory-hard function \cite{OWASP} and then to SHA256 again to shorten the output.\\
The resulting value is a 256-bit (32-ASCII character) value, produced by the final SHA256 hash, and the engineers at Facebook have proudly presented this as being a safe way to upgrade their legacy password storage system without having to create a second password storage table. Thus we presented an overview of the password storage scheme used at Facebook in 2014.

\section{Vulnerability}
In this section, we will present an explanation of a security vulnerability which downgrades the expected security of Facebook's password storage mechanism as described previously. The password storage system at Facebook exists to save costs. It is an ad-hoc solution to an aging information system. Credit must be given to the team who implemented it for their creativity.\\
Unfortunately, this scheme is vulnerable to, at least, a hash collision attack on the initial, broken, MD5 hash algorithm.\\
This is because, at the initial step, the password space is reduced to the MD5 output space, and only from there is it fed into SHA1, then SHA256. This means that all final outputs correspond to an MD5 output, which is not an acceptable security guarantee\cite{CMU,md5_preimage}.
The password storage algorithm detailed previously can be expressed as in \ref{algo}.

\begin{equation}
    \begin{aligned}
    & password &=pw \\ 
    & md5(pw) &=m \\
    & sha1(m,salt) &=s1 \\
    & sha256(s1, secret) &=s2\\
    & scrypt(s2) &=s3 \\
    & sha256(s3) &=value\\
    \end{aligned}
    \label{algo}
\end{equation}

For two different candidate passwords a and b, if $md5(a)=md5(b)$, then with the same salt values for both computations, the final stored value will be identical. See \ref{hack}.

\begin{equation}
    \begin{aligned}
    &&a \not= b \\
    & md5(a) &= m(a) = m(b) \\
    & sha1(m(a),salt(a)) &= s1(a) = s1(b) \\
    & sha256(s1(a), secret) &= s2(a) = s2(b)\\
    & scrypt(s2(a)) &= s3(a) = s3(b) \\
    & sha256(s3(a)) &= value(a) = value(b)\\
    \end{aligned}
    \label{hack}
\end{equation}

We assume the same salt value for a computation because salts are usually stored alongside password hashes in clear text. Their role is not to be secret, but to prevent precomputation attacks \cite{salting}. Furthermore, when attempting to log into a Facebook account secured by password $a$, a wrong candidate password $b$ will be tested using the algorithm with the same salts as applied to $a$. Thus two strings $a$ and $b$ with identical MD5 hashes will be treated by Facebook's authentication system as identical, which is a hash collision attack. Thus we have detailed in this section a security vulnerability which downgrades the expected security of the password storage scheme in use at Facebook to that of the MD5 hashing algorithm with stronger salting and memory-hardness, instead of SHA256 with the same salting and memory-hardness.

\section{Exploit}
In this section we will provide an overview of an exploit which demonstrates the use at Facebook of the insecure password storage scheme detailed previously, and its security weakness. The simplest way to do this is to demonstrate a functional hash collision on Facebook's password storage. We use a pair of strings published on twitter in 2024 by Marc Stevens \cite{tweet}, see \ref{strings}:

\begin{equation}
\begin{aligned}
a = TEXTCOLLBYfGiJUETHQ4h\textcolor{red}{E}cKSMd5zYpgqf1Y\\RDhkmxHkhPWptrkoyz28wnI9V0aHeAuaKnak\\
b = TEXTCOLLBYfGiJUETHQ4h\textcolor{red}{A}cKSMd5zYpgqf1Y\\RDhkmxHkhPWptrkoyz28wnI9V0aHeAuaKnak
\end{aligned}
\label{strings}
\end{equation}

Both of these strings differ by one character and have the same MD5 hash value, see \ref{hash}.

\begin{equation}
    md5(a) = md5(b) = faad49866e9498fc1719f5289e7a0269
    \label{hash}
\end{equation}

Our attack and proof is constructed as follows: we create a Facebook account, for which we set the password to string $a$. We then log out of the account, optionally deleting website data or switching network and device, and we attempt to log back in with string $b$ as the password. At the time of writing, this is systematically successful. We have presented a functional hash collision attack which allows anyone to log into a Facebook account with a candidate password which differs from the actual password if both share the same MD5 hash value.

\section{Consequences}
In the previous section, we have shown a successful hash collision attack which allows someone to bypass Facebook's authentication system with a password candidate that shares the same MD5 hash value as the original password, and we will now discuss the consequences of this exploit. The attack proves that Facebook's password storage is still MD5-dependent, and hence likely the same system that was presented in 2014 \cite{fb_pw_h}. MD5 is a cryptographically broken algorithm and has been since 2008 \cite{CMU}. The output length of MD5 is 128 bits, or 16 ASCII characters in lengths, and the search space, which corresponds to the amount of possible, different user passwords is either 128 or 123 bits \cite{md5_preimage} which is too short by modern standards \cite{CISA_pw}.\\
The consequences of this attack are as follows:
\begin{itemize}
    \item It is is possible to log into a Facebook user account using a string which is not the password.
    
    \item Recovering the password of a leaked Meta or Facebook hash value via exhaustive search guessing requires either $2^{128}$ or $2^{123}$ attempts \cite{md5_preimage}, instead of the expected $2^{256}$. Such leaks are frequent \cite{guardian}.
    
    \item Meta Platforms Inc. does not comply with US NIST standards for the usage of hash functions \cite{SHS}.
    
    \item Meta passwords have a maximal entropy \cite{entropy} of 15 to 16 characters instead of 32. US CISA recommendations are currently of at least 16 characters for all passwords \cite{CISA_pw}, which is either not feasible or the maximum amount at Facebook.
    
    \item The above security issues apply to services and websites which allow users to log in with Facebook via OAuth \cite{OAuth}.
\end{itemize}
In this section, we have detailed the consequences which stem from our attack and demonstration of vulnerability.

\section{Limitations}
In this section, we will attempt to discuss the limitations of our attack.
They are, at least, as follows:
\begin{itemize}
    \item Although we present a login to a Facebook account with an incorrect password, this is possible because we are able to set the account password. Unauthorized access to a Facebook account requires to compute or guess a value which has the same MD5 hash value as the pre-defined password. This is significantly more difficult and likely computationally infeasible for now.
    \item Facebook logins are sometimes secured by 2-factor authentication \cite{2FA} and additional information verifications such as location and device \cite{fb_pw_h}, which prevents unauthorized access.
    \item Due to the researcher's lack of proper knowledge of the best MD5 pre-image attacks, we are unable to determine if the search space of Facebook user passwords is 123 or 128 bits. Both are too short by modern standards but 123 bits would imply that it is not possible to meet current base CISA password recommendations on Facebook's system, \cite{CISA_pw}.
    \item User passwords generally have low entropy. This means that users often have guessable passwords, and that it is not necessary for an attacker to guess all possible password combinations to gain access into an account. Our attack may not be a cost-effective way for an adversary to gain unauthorized access into Facebook accounts.
    \item Facebook engineers have properly secured their system using salts on the SHA256 instance and a memory-hard function which slows password guessing \cite{scrypt_rfc,scrypt2,salting,fb_pw_h}. This implies that it should not be possible for even an extremely powerful attacker to recover the password from a leaked hash for now.
\end{itemize}

\section{Conclusion}
We have provided in this paper an overview of password hashing and of its applications to web and other digital systems account security. We have also detailed and analyzed the password hashing and storage scheme in use at Facebook as presented in 2014. We have identified and explained a vulnerability which severely downgrades the security guarantees of this system, which is used for several billion accounts. We have demonstrated an exploit, still valid at time of writing, which showcases the vulnerability. We have concluded that Facebook's password storage system does not meet current security standards, and that this impacts all applications using Facebook for login purposes. Lastly we have discussed limitations of this exploit, as it does not seem to be immediately usable for unauthorized account takeover.

\section{Disclaimers}
In line with appropriate ethical norms, we have contacted Meta Platforms Inc. about this vulnerability prior to release of public information. Specifically, the researcher has submitted a bug bounty report for this issue on September 19th, 2025. Meta Platforms Inc. Security teams have acknowledged the issue and are working with the relevant product teams who own the password hashing infrastructure to fix the problem. In line with recommendations issued by the CERT at Carnegie Mellon University, which the researcher contacted as the proper vulnerability coordinator for the United States of America, as specified on the webpage of the US Cybersecurity and Infrastructure Security Agency, we are releasing this research 2-3 weeks after acknowledgment by Facebook.

\printbibliography

\end{document}